Т.Г. Третьякова

Новосибирский государственный педагогический университет,

г. Новосибирск

# НЕТРАДИЦИОННАЯ ОБРАБОТКА РЕЗУЛЬТАТОВ ВЫПОЛНЕНИЯ СИ-ТЕСТОВ

0. Данные тезисы подготовлены для доклада на Международной научно-практической конференции "Инновации в педагогическом образовании" (Новосибирский государственный педагогический университет, 22–24 октября 2007 г).

1. Одним из видов интегративных тестов, т.е. тестов, нацеленных на общую диагностику подготовленности [1, с 178], является прагматические тесты, разновидностью которых являются клоуз-тесты, использующиеся в преподавании иностранных языков для определения общего уровня владения языком [3, с. 105]. Ряд недостатков классических клоуз-тестов устранен в Си-тестах, предложенных Раатцем и Клайн-Брелейем. В Си-тестах в каждом втором слове пропущена вторая половина букв. При подсчете результатов учитывается количество точно восстановленных слов, минимальное количество заданий – 100. Трудность Си-теста определяется трудностью исходного неискаженного теста.

Несмотря на традиционное использование в названии контролирующего материала слова *тест*, Си-тесты тестами не являются, поскольку в них нарушается свойство локальной независимости тестовых заданий, согласно которому результат выполнения одного задания не зависит от результата выполнения другого [4, с. 314]. (Си-тест составлен на основе единого текста с общей тематикой, и правильное заполнение одного пропуска способствует заполнению связанных с этим местом текста пропусков).

2. В традиционных педагогических тестах на этапе чистки теста измеряются интеркорреляции между заданиями (коэффициенты корреляции Пирсона между вектор-столбцами заданий), и из двух тесно связанных заданий одно удаляется. Ввиду большой практической ценности и популярности Си-тестов нами предлагается методика обработки результатов тестирования Си-тестами, позволяющая считать Си-тесты контролирующими материалами с выполнением гипотезы локальной независимости, а тем самым повысить валидность тестового результата.

Результат выполнения теста по определенному заданию характеризуется вектором задания, составленным из нулей и единиц в зависимости от успешности выполнения задания испытуемыми. Назовем расстоянием между двумя заданиями количество несовпадений между векторами, нормированное на единицу. Иллюстративный пример: тест выполняется 10 испытуемыми. Задание 1 характеризуется вектором (1,1,0,1,1,0,0,0,0,0), задание 2 – вектором (1,0,1,1,1,0,1,0,0,0). Расстояние между заданиями составляет $(0 + 1 + 1 + 0 + 0 + 0 + 1 + 0 + 0 + 0)/10 = 0{,}3$. Ввиду известных недостатков коэффициента корреляции Пирсона как меры взаимосвязи, предлагается оценивать интеркорреляции между заданиями введенной мерой расстояния между ними, и группировать задания в кластеры на основе близости между ними (к одному кластеру относятся задания с расстоянием $a < a_{\text{крит}}$, где $a_{\text{крит}}$ – некое критическое значение расстояния между заданиями. Кластеризация по выдвинутому признаку носит не жестко детерминированный, а вероятностный характер, т.е. принадлежность заданий А, Б и В к одному кластеру на основании расстояний АБ и АВ не означает обязательности принадлежности заданий Б и В к одному кластеру на основании расстояния БВ (хотя подобное и не исключено).

Гипотеза исследования заключается в том, что при приписывании заданиям (пропускам) Си-теста переменных весов, обратно пропорциональных количеству заданий в соответствующем кластере, тестовые результаты будут более адекватно отражать измеряемое свойство или комплекс свойств. Гипотеза базируется на утверждении, что при оценке объединенных в кластер заданий общий балл за выполнение всех заданий кластера составит 1, т.е. появляется возможность рассматривать Си-тест как традиционный тест с количеством заданий, равных количеству созданных кластеров, и низкой интеркорреляцией между заданиями-кластерами.

3. Для проверки гипотезы использованы результаты тестирования студентов 4 курса ФИЯ НГПУ Си-тестом на основе учебного текста *Als ich das erste Mal auf Deutsch träumte*. Си-тест содержит 145 пропусков-заданий, количество испытуемых составляло 54 человека. При традиционной оценке, когда каждый верно восстановленный пропуск оценивался 1 баллом, средний тестовый балл составил 84,2 при стандартном отклонении 20,3, что дает коэффициент вариации $v = 20,3/84,2 = 0,241$. Проведен ряд обработок результатов с изменением величины $a_{крит}$ с целью достижения наилучшей вариативности результата. Известно, что вариативность результата кладется в основу определения оптимального времени тестирования (2, с. 134), и мы считаем правомерным использование этого-же принципа для определения наиболее оптимального значения $a_{крит}$. Наилучший результат достигнут при $a_{крит} = 0,25$ – среднее по тесту составило 22,6 при стандартном отклонении 8,0 и коэффициенте вариации 0,352. При $a_{крит} = 0,25$ сумма весов всех заданий составляет 44,4, т.е. в среднем в один кластер (измеряющий какое-либо одно подсвойство в пределах измеряемого Си-тестом свойства) попадают $145/44,4 = 3,26$ задания. В то-же время в Си-тесте выделяются 26 кластеров, состоящих из одного пропуска-задания, которым присваивается максимально возможный вес. Анализ текста показывает.

что это высоковалидные задания, которыми с высокой надежностью проверяется наличие или отсутствие измеряемого свойства (комплекса свойств).

В один кластер в соответствии с предложенным принципом кластеризации объединяются задания чересчур легкие и чересчур трудные. Тем самым эти задания оцениваются с низким весом и практически исключаются из теста, что отвечает общему принципу включения в традиционный нормативно-ориентированный тест лишь заданий с коэффициентом выполнения от 30–35% до 80–85% [3, с. 59].

Предварительные результаты исследования возможности применения предложенного метода обработки результатов тестирования студентов, изучающих иностранный язык, Си-тестом, свидетельствуют в пользу метода вследствие повышения вариативности результата. Поскольку тестовые результаты перестают зависеть от особенностей исходного положенного в основу Си-теста текста (увеличение доли заданий, проверяющих одно подсвойство, не увеличивает вес этого подсвойства в общем результате), можно говорить об увеличении валидности результата, который напрямую связывается с количеством освоенных испытуемым подсвойств.

**Список литературы**